\documentclass[fleqn,usenatbib]{mnras}
\usepackage{amsmath}
\usepackage{xspace}
\usepackage{mathptmx}
\usepackage{txfonts}
\usepackage{xcolor}
\usepackage[T1]{fontenc}
\usepackage{ae,aecompl}
\usepackage{float}

\usepackage{graphicx}	
\usepackage{amssymb}	
\usepackage{booktabs}
\usepackage{wasysym}

\let\oldhref\href
\renewcommand{\href}[2]{\oldhref{#1}{\hbox{#2}}}

\definecolor{colorl1}{RGB}{0, 51, 153}
\definecolor{colorl2}{RGB}{153, 0, 0}
\definecolor{colorl3}{RGB}{179, 179, 0}
\definecolor{colorl4}{RGB}{51, 102, 0}

\definecolor{colorw1}{RGB}{51, 102, 255}
\definecolor{colorw2}{RGB}{255, 51, 0}
\definecolor{colorw3}{RGB}{255, 214, 51}
\definecolor{colorw4}{RGB}{51, 204, 51}

\newcommand{\hMpc}{{\ifmmode{h^{-1}{\rm Mpc}}\else{$h^{-1}$Mpc}\fi}}
\newcommand{\Mpc}{{\ifmmode{{\rm Mpc}}\else{Mpc}\fi}}
\newcommand{\hkpc}{{\ifmmode{h^{-1}{\rm kpc}}\else{$h^{-1}$kpc}\fi}}
\newcommand{\kpc}{{\ifmmode{ {\rm kpc} }\else{{\rm kpc}}\fi}}
\newcommand{\kms}{{\ifmmode{ {\rm km\,s^{-1}} }\else{ ${\rm km\,s^{-1}}$ }\fi}}
\newcommand{\hMsun}{{\ifmmode{h^{-1}{\rm {M_{\astrosun}}}}\else{$h^{-1}{\rm{M_{\astrosun}}}$}\fi}}
\newcommand{\Msun}{{\ifmmode{{\rm M}_{\astrosun}}\else{${\rm M}_{\astrosun}$}\fi}}
                        
\newcommand{\Mhalo}{{\ifmmode{M_{\rm halo}}\else{$M_{\rm halo}$}\fi}}
\newcommand{\Rvir}{{\ifmmode{R_{\rm vir}}\else{$R_{\rm vir}$}\fi}}
\newcommand{\Mvir}{{\ifmmode{M_{\rm vir}}\else{$M_{\rm vir}$}\fi}}
\newcommand{\Mstar}{{\ifmmode{M_{\rm star}}\else{$M_{\rm star}$}\fi}}
\newcommand{\Vrot}{{\ifmmode{V_{\rm rot}}\else{$V_{\rm rot}$}\fi}}
\newcommand{\ltsima}{$\; \buildrel < \over \sim \;$}
\newcommand{\gtsima}{$\; \buildrel > \over \sim \;$}
\newcommand{\lsim}{\lower.5ex\hbox{\ltsima}}
\newcommand{\gsim}{\lower.5ex\hbox{\gtsima}}

\def\lesssim{\mathrel{\hbox{\rlap{\hbox{\lower4pt\hbox{$\sim$}}}\hbox{$<$}}}}
\def\gtrsim{\mathrel{\hbox{\rlap{\hbox{\lower4pt\hbox{$\sim$}}}\hbox{$>$}}}}

\newcommand{\beq}{\begin{equation}}
\newcommand{\eeq}{\end{equation}}
\def\beqa{\begin{eqnarray}}
\def\eeqa{\end{eqnarray}}
\def\LCDM{\ensuremath{\Lambda}CDM}

\def\head{ \vbox to 0pt{\vss \hbox to 0pt{\hskip 440pt\rm
      LA-UR-10-07069\hss} \vskip 25pt}}

\def \kms {\ifmmode  \,\rm km\,s^{-1} \else $\,\rm km\,s^{-1}  $ \fi }
\def \kpc {\ifmmode  {\,\rm kpc}  \else ${\rm  kpc}$ \fi  }  
\def \hkpc {\ifmmode  {h^{-1}\rm kpc}  \else ${h^{-1}\rm kpc}$ \fi  }  
\def \hMpc {\ifmmode  {h^{-1}\rm Mpc}  \else ${h^{-1}\rm Mpc}$ \fi  }  
\def \Mpch {\ifmmode  {h^{-1}\rm Mpc}  \else ${h^{-1}\rm Mpc}$ \fi  }  
\def \Msun {\ifmmode {\rm M}_{\astrosun} \else ${\rm M}_{\astrosun}$ \fi} 
\def \hMsun {\ifmmode h^{-1}\,\rm M_{\astrosun} \else $h^{-1}\,\rm M_{\astrosun}$ \fi}
\def \Gyr {\ifmmode\, \rm Gyr \else $\,$Gyr \fi}

\def \LCDM {\ifmmode \Lambda{\rm CDM} \else $\Lambda{\rm CDM}$ \fi}
\def \OmegaM {\ifmmode \Omega_{\rm m} \else $\Omega_{\rm m}$ \fi} 
\def \Omegab {\ifmmode \Omega_{\rm b} \else $\Omega_{\rm b}$ \fi} 
\def \OmegaL {\ifmmode \Omega_{\rm \Lambda} \else $\Omega_{\rm \Lambda}$\fi} 
\def \Deltavir {\ifmmode \Delta_{\rm vir} \else $\Delta_{\rm vir}$ \fi}
\def \rhocrit {\ifmmode \rho_{\rm crit} \else $\rho_{\rm crit}$ \fi}
\def \rhou {\ifmmode \rho_{\rm u} \else $\rho_{\rm u}$ \fi}
\def \zc {\ifmmode z_{\rm c} \else $z_{\rm c}$ \fi}

\title[BH Accretion] {Co-Evolution vs. Co-existence: The Effect of Accretion Modelling on the Evolution of Black Holes and Host Galaxies}

\author[N.H. Soliman et al.]{Nadine H. Soliman$^{1}$\thanks{E-mail: nsoliman@caltech.edu}, Andrea V. Macci\`o$^{2,3,4}$, Marvin Blank$^{2,3,5}$
\\
$^{1}$TAPIR, Mailcode 350-17, California Institute of Technology, Pasadena, CA 91125, USA \\
$^{2}$New York University Abu Dhabi, PO Box 129188, Abu Dhabi, United Arab Emirates \\
$^3$Center for Astro, Particle and Planetary Physics (CAP$^3$), New York University Abu Dhabi\\
$^4$Max-Planck-Institut f\"ur Astronomie, K\"onigstuhl 17, 69117 Heidelberg, Germany \\
$^5$Institut f\"{u}r Theoretische Physik und Astrophysik, Christian-Albrechts-Universit\"{a}t zu Kiel, Leibnizstr. 15, D-24118 Kiel, Germany\\}

\date{Accepted XXX. Received YYY; in original form ZZZ}

\pubyear{2023}

\begin{document}

\label{firstpage}
\pagerange{\pageref{firstpage}--\pageref{lastpage}}
\maketitle
\begin{abstract}
We append two additional black hole (BH) accretion models, namely viscous disc and gravitational torque-driven accretion, into the Numerical Investigation of a Hundred Astrophysical Objects (NIHAO) project of galaxy simulations. We show that these accretion models, characterized by a weaker dependence on the BH mass compared to the commonly used Bondi-Hoyle accretion, naturally create a common evolutionary track (co-existence) between the mass of the BH and the stellar mass of the galaxy, even without any direct coupling via feedback (FB). While FB is indeed required to control the final BH and stellar mass of the galaxies, our results suggest that FB might not be the leading driver of the cosmic co-evolution between these two quantities; in these models, co-evolution is simply determined by the shared central gas supply. Conversely, simulations using Bondi-Hoyle accretion show a two-step evolution, with an early growth of stellar mass followed by exponential growth of the central supermassive black hole (SMBH). Our results show that the modelling of BH accretion (sometimes overlooked) is an extremely important part of BH evolution and can improve our understanding of how scaling relations emerge and evolve, and whether SMBH and stellar mass co-exist or co-evolve through cosmic time.

\end{abstract}

\begin{keywords}
 galaxies: evolution -- 
 galaxies: active --
 galaxies: formation --
 galaxies: nuclei -- methods: numerical -- quasars:general
\end{keywords}

\section{Introduction}

Black holes (BH) are of specific interest in the field of cosmology as they are abundant objects, with almost all known galaxies hosting supermassive black holes (SMBH) at their centre \citep{kormendy1995inward, magorrian1998demography}. Further, observational studies have demonstrated multiple correlations between the BH mass and the host galaxy’s stellar properties (i.e., $M_{\rm BH}$–$M_{\rm bulge}$, $M_{\rm BH}$–$\sigma$ relations; \citep{magorrian1998demography, ferrarese2000fundamental, gebhardt2000relationship, haring2004black, gultekin2009m,sani2011spitzer, kormendy_ho13, mcconnell_etal13, shankar2016selection}). 

These tight correlations present an important question regarding the role SMBHs play in their host’s evolution. Their origin could be interpreted in two ways. First, they could be the result of mere co-existence implying a non-causal effect, i.e., hierarchical merging, or a physical local link between BH growth and star formation as the processes compete for a common gas supply \citep{Peng_2007, jahnke2011non, silk1998quasars, fabian1999obscured}.
On the other hand, the co-evolution hypothesis proposes a direct influence of BHs on their hosts, implying that BHs can affect galaxy-wide processes to a greater extent than the previous interpretation by coupling with the galaxy's environment. Traditionally, the mechanism invoked is Active Galactic Nuclei (AGN) feedback (FB), whereas the BH accretes gas, it releases energy and/or momentum back into the system altering the state of the surrounding gas making it unavailable for star formation \citep{silk1998quasars, fabian1999obscured, king03, begelman_nath05, murray05, mcquillin2012momentum}. During periods of high BH accretion, the FB provides large amounts of energy which substantially affects global processes within the host galaxy resulting in BH-host galaxy co-evolution \citep{fabian2012observational}. Thus in this interpretation, the coupling of the energy emitted from the BH and its surroundings gives rise to a BH that is self-regulating its growth and co-evolving with its host. This mechanism has gained wide support as it could provide large-scale outflows and heat the molecular gas within the galaxy, thus providing several pathways to prevent the over-production of stars in high-mass galaxies \citep{Springel2005, di2008direct, Power06, booth2009cosmological, debuhr2011, kimm_etal11, Choi_2012}.

In addition, AGN activity has been correlated with red sequence galaxies \citep{schawinski2009moderate, silverman2008evolution, Hardcastle2020}, suggesting a connection between FB and galaxy properties. However, while strong (quasar) AGN activity is frequently observed at high redshift \citep{nesvadba2008evidence}, there is no direct evidence linking it to the quenching of star formation. For instance, \citet{mendel2013towards} find no excess optical AGN activity in recently quenched galaxies. Nonetheless, simulations have demonstrated the importance of AGN FB in successfully quenching star formation in massive galaxies \citep{Schaye2015, springel2005modelling, hopkins2006fueling}. These simulations often adopt the Bondi-Hoyle accretion prescription, where FB is necessary to regulate the accretion rates for massive BHs.

 However, alternative BH accretion prescriptions have been proposed in the literature. For instance, \citet{angles2013black} argue that the gravitational torque-based accretion model introduced by \citet{hopkins2011analytic} does not require FB to regulate BH growth or reproduce scaling relations. They propose that the accretion rate is limited by the inflow of material onto the galactic disc, and find that the galaxies converge onto the scaling relations without including AGN FB. Similarly, \citet{debuhr2011} suggest that when utilizing an accretion prescription based on viscous disc ($\alpha$-disc) accretion, BH growth can be self-regulated through FB. However, they also argue that achieving the scaling relations does not necessarily require the involvement of AGN FB to suppress star formation. In other words, while BH growth may be self-regulated, its dominant role in regulating star formation activity in galaxies may not be necessary.

It is worth noting that these simulations utilize different physics, resolutions, and FB implementations, making direct comparisons challenging. Thus, to address whether SMBHs co-exist or co-evolve with their host's stellar mass, self-consistent studies incorporating various models of BH accretion within a single simulation suite are needed. In this paper, we implement two sub-grid models of BH accretion, namely, gravitational torque-based accretion \citep{hopkins2011analytic} and viscous disc-based accretion \citep{shakura1973black, debuhr2011}, in addition to the Bondi-Hoyle accretion model introduced in \citet{Blank2019} to the NIHAO simulation suite. 

\section{Simulations} \label{sec:simulations}

This work is based on a subset of galaxies selected from the  NIHAO (Numerical Investigation of Hundred Astrophysical Objects) suite of cosmological hydrodynamical simulations \citep{Wang2015,Blank2019}

The NIHAO simulations are based on the {\sc gasoline2} code \citep{Wadsley2017},
and include Compton cooling and photoionisation and heating from the ultraviolet background following \citet{Haardt2012}, metal cooling, chemical enrichment, star formation and FB from supernovae and massive stars \citep[the so-called Early Stellar Feedback][]{Stinson13}.
The cosmological parameters are set according to \cite{Planck2014}: Hubble parameter
$H_0$= 67.1 \kms Mpc$^{-1}$, matter density $\Omega_\mathrm{m}=0.3175$, dark energy density
$\Omega_{\Lambda}=1-\Omega_\mathrm{m} -\Omega_\mathrm{r}=0.6824$, baryon density
$\Omega_\mathrm{b}=0.0490$, normalization of the power spectrum $\sigma_8 = 0.8344$, 
slope of the initial power spectrum $n=0.9624$, and each galaxy is resolved with at least
half a million elements (dark matter, gas and stars). The mass and spatial resolution vary across the whole sample, from a dark matter particle mass of 
$m_{\rm dm}  = 3.4 \times 10^3$ \Msun (and force softening of $\epsilon =100$  pc) for dwarf galaxies to $m_{\rm dm} = 1.4 \times 10^7$ \Msun and $\epsilon = 1.8$ kpc for the most massive galaxies \citep[see][for more details]{Wang2015,Blank2019}.
The NIHAO simulations have been proven to be very successful in reproducing several observed
scaling relations like the Stellar Halo-Mass relation \citep{Wang2015}, the disc gas mass and disc size relation \citep{Maccio2016}, the Tully-Fisher relation \citep{Dutton2017}, the star formation main sequence \citep{blank2021nihao}, and the $M_{*}-M_{\rm BH}$ relation \citep{Blank2019}. 

For this work, we concentrate our attention on zoom-in simulations with major merger events of four galaxies: g1.55e12, g2.37e12, g2.71e12 and g6.86e12\footnote{In the NIHAO suite, the name of a galaxy indicates its total virial mass in the low-resolution run.}. The properties of these galaxies in the original NIHAO runs and for different accretion models presented in this paper are reported in Table \ref{tab:sims}.

In the next section, we summarize the main features of BH accretion and FB used in the standard NIHAO simulations, together with the newly implemented
accretion modes for this study. To make the comparison between the effects of different accretion models easier, all simulations run with FB employ the same energy deposition model, with no re-calibration of its parameters.

\begin{table*}
\centering
\begin{tabular}{llcccccc}
\hline   
Galaxy & Model  & $m_{\rm dm} [\Msun]$ & $m_{\rm b} [\Msun]$  &$M_{\rm vir} [\Msun] $ & $M_*$ [\Msun] & $M_{\rm BH} [\Msun]$\\
\hline
\hline
      g1.55e12 & Bondi-Hoyle & $1.7 \times 10^6$ &$6.3 \times 10^4$ & 1.1 $\times 10^{12}$ & 4.1 $\times 10^{10}$ & 0.7  $\times 10^8$  \\
      g1.55e12 & Torque & $1.7 \times 10^6$ &$6.3 \times 10^4$& 1.3 $\times 10^{12}$ & 8.1 $\times 10^{10}$ & 4.3  $\times 10^9$  \\
      g1.55e12 &Torque+FB & $1.7 \times 10^6$ &$6.3 \times 10^4$&  1.0 $\times 10^{12}$ & 2.5 $\times 10^{10}$ & 1.7  $\times 10^8$  \\
     g1.55e12 &$\alpha$-disc &$1.7 \times 10^6$ & $6.3 \times 10^4$& 1.3 $\times 10^{12}$ & 8.0 $\times 10^{10}$ & 1.7  $\times 10^9$  \\
    g1.55e12 &  $\alpha$-disc+FB&$1.7 \times 10^6$ & $6.3 \times 10^4$& 0.9 $\times 10^{12}$ & 0.4 $\times 10^{10}$ & 0.5  $\times 10^8$  \\
\hline
      g2.37e12 & Bondi-Hoyle & $1.7 \times 10^6$ & $6.3 \times 10^4$& 2.0 $\times 10^{12}$ & 5.4 $\times 10^{10}$ & 2.0  $\times 10^8$  \\
      g2.37e12 & Torque & $1.7 \times 10^6$ & $6.3 \times 10^4$& 2.3 $\times 10^{12}$ & 1.9 $\times 10^{11}$ & 1.9  $\times 10^9$  \\
      g2.37e12 &  Torque+FB &$1.7 \times 10^6$ & $6.3 \times 10^4$ & 2.1 $\times 10^{12}$ & 7.4 $\times 10^{10}$ & 8.0  $\times 10^8$  \\
     g2.37e12 &  $\alpha$-disc &$1.7 \times 10^6$ & $6.3 \times 10^4$& 2.3 $\times 10^{12}$ & 1.8 $\times 10^{11}$ & 1.9  $\times 10^9$  \\
 g2.37e12 &  $\alpha$-disc+FB& $1.7 \times 10^6$ & $6.3 \times 10^4$&1.9 $\times 10^{12}$ & 1.2 $\times 10^{10}$ & 1.8  $\times 10^8$  \\
\hline
 g2.71e12 & Bondi-Hoyle & $1.7 \times 10^6$ & $6.3 \times 10^4$ & 2.2 $\times 10^{12}$ & 3.5 $\times 10^{10}$ & 1.6  $\times 10^8$  \\
      g2.71e12 & Torque & $1.7 \times 10^6$ & $6.3 \times 10^4$& 2.5 $\times 10^{12}$ & 1.5 $\times 10^{11}$ & 5.3  $\times 10^8$  \\
      g2.71e12 & Torque+FB & $1.7 \times 10^6$ & $6.3 \times 10^4$& 2.3 $\times 10^{12}$ & 7.8 $\times 10^{10}$ & 5.6  $\times 10^8$  \\
     g2.71e12 & $\alpha$-disc&  $1.7 \times 10^6$ & $6.3 \times 10^4$& 2.6 $\times 10^{12}$ & 1.5 $\times 10^{11}$ & 1.8  $\times 10^9$  \\
 g2.71e12 &  $\alpha$-disc+FB & $1.7 \times 10^6$ & $6.3 \times 10^4$& 2.0 $\times 10^{12}$ & 1.0 $\times 10^{10}$ & 1.54  $\times 10^8$  \\
\hline
      g6.86e12 & Torque (1 kpc) & $1.7 \times 10^6$ & $1.5 \times 10^4$&  4.7  $\times 10^{12}$&  1.8  $\times 10^{11}$&  1.8  $\times 10^{9}$\\
      g6.86e12 &  Torque (0.5 kpc)& $1.7 \times 10^6$ & $1.5 \times 10^4$& 4.5  $\times 10^{12}$&  1.9  $\times 10^{11}$&  6.4 $\times 10^{8}$\\
     g6.86e12 & $\alpha$-disc  (1 kpc)& $1.7 \times 10^6$ & $1.5 \times 10^4$&  4.8 $\times 10^{12}$ &  1.8 $\times 10^{11}$ &  7.7 $\times 10^{8}$\\
    g6.86e12 &$\alpha$-disc (0.5 kpc) & $1.7 \times 10^6$ & $1.5 \times 10^4$&  4.5  $\times 10^{12}$&  1.8 $\times 10^{11}$ &  3.0 $\times 10^{8}$ \\
     g6.86e12 & $\alpha$-disc (0.25 kpc)& $1.7 \times 10^6$ & $1.5 \times 10^4$&  4.7 $\times 10^{12}$ &  1.8  $\times 10^{11}$&  0.9 $\times 10^{8}$ \\

\hline
\hline
\end{tabular}
    \caption{Galaxy parameters at $z=0$ for different black hole (BH}) accretion models with and without FB for our fiducial accretion parameter runs unless otherwise stated. The Bondi-Hoyle model has been only run with FB. The columns show: (1) Galaxy Name. (2) BH accretion model. (3) Dark matter particle mass: $m_{\rm dm}$. (4) Baryonic matter Particle mass: $m_{\rm b}$. (5) Virial mass: $M_{\rm vir}$. (6) Stellar mass: $M_*$. (7) BH Mass: $M_{\rm BH}$.For galaxy g6.86e12, simulations were run until redshift $z \sim 1$ and the accretion radius used for each model is provided in parentheses.
    \label{tab:sims}
\end{table*}

\section{Computational Methods for Black Hole Physics} \label{sec:methods}
The physical processes that govern the growth of SMBHs inherently span a wide range of physical scales. This makes modelling BH growth and incorporating accurately evolving SMBHs in simulations of galaxy formation a challenging task. Due to constraints imposed by computational resources and limited resolution, sub-grid models are employed. These models parameterize the effects of small-scale unresolved processes, such as those occurring on accretion disc scales, to approximate an effective accretion rate. In the following subsections, we describe our implementation of sub-grid prescriptions of the Bondi-Hoyle, gravitational torque, and viscous disc accretion models.

In this study, we model BHs as sink particles that interact with their environment only through gravitational forces and accrete surrounding gas particles. 

In our simulations, BHs are seeded with an initial mass of $10^5 \, \Msun$ into all central halos with masses greater than $5\times10^{10}\, \Msun$. During the simulation, BHs are allowed to merge when their distance is smaller than the sum of their softening lengths. To address the issue of unnatural wandering of BHs within the simulation, we have adopted a BH relocation scheme. This scheme involves setting the position and velocity of the BH to match those of a dark matter particle with the lowest gravitational potential within a radius of 10 softening lengths. By adopting this approach, we effectively prevent the BH from drifting unnaturally and ensure its alignment with the dominant gravitational structures in its vicinity. For more details on the exact implementation of seeding, merging, and relocation, refer to \citet{Blank2019}.

\subsection{Shared Features For All Accretions Models}
The BH accretion rate predicted by each of the models is capped at the Eddington rate \citep{eddington1921strahlungsgleichgewicht},

\begin{equation}
    \dot{M}_{\rm Edd} = \frac{M_{\text{BH}}}{\epsilon_{\rm r}\tau_{\rm s}},
    \label{eq:edd}
\end{equation}
with the Salpeter time-scale $\tau_{\rm S} = 4.5 \times 10^8$ yr \citep{salpeter1964accretion} and the radiative efficiency $\epsilon_{\rm r} = 0.1$ \citep{shakura1973black}. Thus, we set the BH accretion rate as the minimum between the accretion rate computed directly from the code (for each of the different models) and the one given by Equation \ref{eq:edd}. 

For a given value of the accretion rate, the most gravitationally bound gas particle(s), or fractions of particles, are accreted onto the BH. For each time step $\Delta t$, the BH accretes mass $\Delta t \,\dot{M}_{\text{BH}}$, and the momentum of the gas particles is then added to the BH's momentum. As fractions of a gas particle's mass could be accreted, we remove gas particles with a mass less than 20\% of their initial mass; their momentum and mass are then redistributed to the surrounding gas particles weighted by the SPH kernel. As the mass of the BH grows, its softening length is multiplied by $(1 + \Delta M/M_{\text{BH}})^{1/2}$ for an increase of $\Delta M$ in mass.  

\subsection{Feedback Model}
For our simulations that incorporate AGN FB, we couple the AGN to the surrounding gas through a spherical thermal input FB model. In this model, the AGN luminosity associated with an accretion rate of $\dot{M}_{\rm BH}$ is given by
\begin{equation}
    L = \epsilon_{\rm r} \dot{M}_{\text{BH}} c^2,
\end{equation}
with $c$ being the speed of light. 
A fraction $\epsilon_{\rm f}$ = $0.05$ of the luminosity is then given as thermal energy ($\dot E$) to the nearest 50 particles, weighted by the SPH kernel:
\begin{equation}
    \label{eq:fb}
    \dot{E} = \epsilon_{\rm f} \epsilon_{\rm r} \dot{M}_{\rm BH} c^2.
\end{equation}
To avoid too large sound speeds and thus too small time-steps, the specific energy of a single gas particle is limited to $(0.1c)^2$, which roughly corresponds to $\sim 10^7$K.

\subsection{Bondi-Hoyle Accretion}
 Bondi-Hoyle accretion, a model introduced in \citet{ hoyle1939effect, bondi1944mechanism, bondi1952spherically}, was formulated to describe the accretion rate of a point mass moving through a uniform spherical gas cloud. We utilize a Bondi-Hoyle-like sub-grid prescription that is widely adopted in the literature \citep[e.g., ][]{di2005energy, springel2005modelling,di2008direct,booth2009cosmological, Choi_2012, johansson2009evolution, croft2009galaxy, colberg2008supermassive}.  As shown in Equation \ref{eq1}, the model describes an accretion rate for a BH of mass $M_{\rm BH}$, moving at velocity $v$ relative to its surrounding gas of density $\rho$ and speed of sound $c_{\rm s}$. The accretion rate is boosted by the dimensionless parameter $\alpha$, proposed by \cite{springel2005modelling} to correct for the likely underestimated values of $\rho$ due to the effects of coarse resolution. We adopt a constant value of $\alpha=70$ throughout our simulations, based on the parameter study shown in \citet{Blank2019}.
\begin{equation}
    \label{eq1}
    \dot{M}_{\rm Bondi} = \frac{4 \pi \alpha G^2 M^2_{\rm BH} \rho}{(c_{\rm s}^2 + v^2)^{3/2}}.
\end{equation} 
We evaluate the parameters $\rho$, $c_{\rm s}$ and $v$  as the kernel weighted averages evaluated at the location of the BH within a radius that encompasses the nearest 50 gas particles. This model is characterised by an accretion rate that strongly depends on the mass of the BH, i.e. $\dot{M}_{\rm Bondi} \propto M_{\rm BH}^2$. Therefore, this model requires the inclusion of some form of AGN FB to limit gas inflow as the BH increases in mass and prevents BHs from growing to unrealistic masses. In this paper, we always run Bondi-Hoyle accretion with FB.

\subsection{Gravitational Torque-driven Accretion}
 The gravitational torque sub-grid model, proposed by \citet{hopkins2011analytic}, is an analytical model that is physically motivated by considering BH accretion due to the inflow of material from galactic disc scales down to unresolved sub-parsec scales.
 
As shown in Equation \ref{eq3} below, the model predicts that accretion rates increase linearly with the mass of the galactic disc $M_{\rm disc}$ and are strongly influenced by the fraction of stellar and gaseous matter in the disc $f_{\rm disc}$. However, the dependence on the mass of the BH $M_{\rm BH}$ is relatively weak. The accretion rate can be expressed as follows:
\begin{equation}
    \label{eq3}
    \begin{split}
    \dot{M}_{\text{Torque}} \approx  \epsilon_\text{T} f_{\text{disc}}^{5/2} \;
    & \bigg(\frac{M_{\text{BH}}}{10^8 M_\odot}\bigg)^{1/6} \bigg(\frac{M_{\text{disc}}(R_0)}{10^9 M_\odot}\bigg) \\
    & \bigg(\frac{R_0}{100 \text{pc}}\bigg)^{-3/2}\bigg(1 + \frac{f_0}{f_{\text{gas}}}\bigg)^{-1} \; M_\odot \: \text{yr}^{-1}, 
    \end{split}
\end{equation}
where
\begin{equation}
    f_0 \approx 0.31 f^2_{\text{disc}}\big(M_{\text{disc}}(R_0)/10^9 M_\odot \big)^{-1/3}, 
\end{equation}
\begin{equation}
    f_{\text{disc}}=  M_{\text{disc}}/\big(M_{\text{gas}}(R_0) + M_{\text{star}}(R_0)\big), 
\end{equation}
\begin{equation}
    f_{\text{gas}}(R_0)\equiv M_{\text{gas}}(R_0)/M_{\text{disc}}(R_0), 
\end{equation}
  where $M_{\text{star}}$ and $M_{\text{gas}}$ correspond to the total stellar and gas masses respectively, and $M_{\text{disc}}$ is the total mass of stars and gas located in the galactic disc. These parameters are evaluated within a radius $R_0$, which for this model is defined as the radius that encompasses the closest 256 gas particles closest to the BH, with an upper limit of 1 kpc unless otherwise specified. To estimate $M_{\text{disc}}$, we use the kinematic decomposition scheme outlined in \cite{angles2013black}. In our evaluations, we highlight that the radius $R_0$ encloses a larger number of particles compared to our Bondi-Hoyle or viscous disc accretion rate calculations. This discrepancy arises from the need to employ a greater number of particles for the kinematic decomposition process. We assume a normalization factor $\epsilon_\text{T}=0.5$ as calibrated in \citet{angles2016}, which encapsulates the efficiency of gaseous transport from the galactic disc onto the BH. To ensure that the calibration produces realistic BH masses in our simulations, we conduct post-processing analyses that show agreement with scaling relations.

The effective total accretion rate is calculated using Equation \ref{eq4} shown below \citep{yuconst}, 
\begin{equation}
    \label{eq4}
    \dot{M}_{\text{BH}} = (1- \epsilon_{\rm r})  \, \dot{M}_{\text{Torque}}.
\end{equation}

Unlike the Bondi-Hoyle model, the accretion rate predicted by this model has a weak dependence on the mass of the BH,  $\dot{M}_{\text{Torque}} \propto M_{\text{BH}}^{1/6}$ and a higher dependence on the local environment surrounding the BH such as $f_{\text{disc}}$ and $M_{\text{disc}}$. Thus, this model is less susceptible to runaway effects (in contrast to the Bondi-Hoyle model) and does not rely heavily on AGN FB to limit the growth of the BH at later times or to match observational scaling relations. 

\subsection{Viscous Disc Accretion}
\label{sec:visc}

The viscous disc model presented in \citet{debuhr2011} is a sub-grid model that estimates accretion due to angular momentum dissipation by viscous forces on unresolved scales.
In this model, the accretion rate within a given radius $R_0$ depends on the surface density of the gas $\Sigma$, the sound speed $c_{\rm s}$, and the orbital velocity of the gas $\Omega$, as shown in the following equation: 

\begin{equation}
    \label{eq6}
    \dot{M}_{\alpha-\text{disc}} = 3 \pi \alpha \Sigma \frac{c_{\rm s}^2}{\Omega},
\end{equation}
where
\begin{equation}
    \Sigma = M_{\rm gas}(R_0)/(\pi R^2_0),
\end{equation}
and
\begin{equation}
    \Omega^2 \approx GM(R_0)/(R^3_0),
\end{equation}
with $M_{\rm gas}(R_0)$, and $M(R_0)$ being the total gas mass and the total mass within $R_0$ respectively. We calculate all the quantities mentioned above within a radius $R_0$. For this model, $R_0$ corresponds to the radius that encompasses the 50 nearest gas particles, with an upper limit of 1 kpc, unless otherwise specified. The parameter $\alpha$, similar to the $\alpha$-disc model introduced in \citet{shakura1973black}, approximates the efficiency of angular momentum transport and encapsulates the effects of processes that take place on unresolved scales. For our simulations, we normalize $\alpha = 1.4 \times 10^{-3}$, calibrated such that our galaxies (simulated without FB) are consistent with the observed Stellar Mass-Halo Mass and the BH Mass-Stellar Mass relations. This model does not include a direct dependence on the BH mass (apart from its contribution to $M(<R_0)$), and only depends on the local environment. Furthermore, the model does not rely on AGN FB to reproduce the scaling relations as demonstrated in \citet{debuhr2011}. 

\section{Results}\label{sec:results}

\subsection{Evolution of Star Formation Rates}
\begin{figure}[H]
    \centering
    \includegraphics[width = 0.45\textwidth]{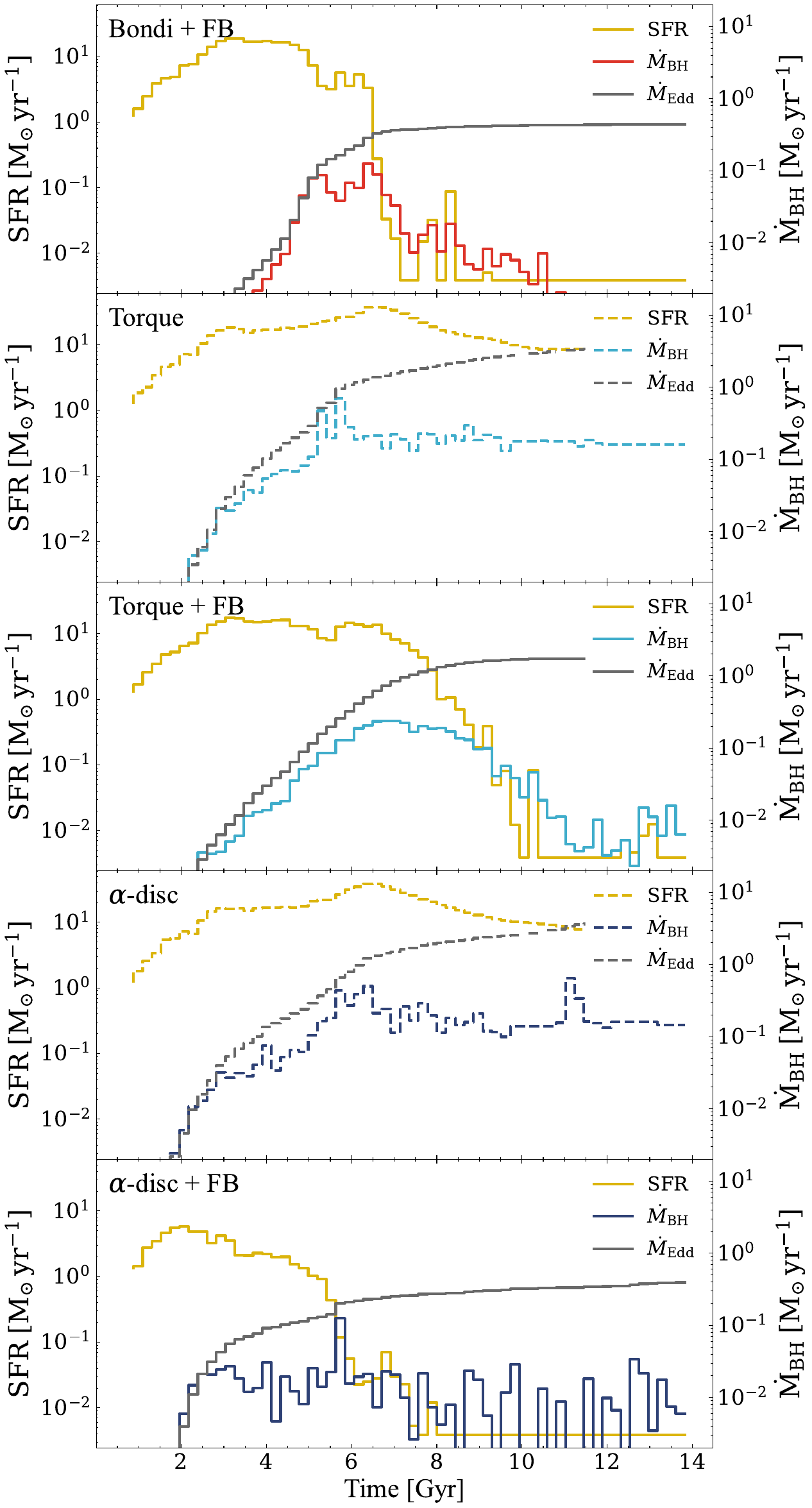}
    \caption{From top to bottom: Star formation and black hole (BH) accretion rates of the central BH as a function of time for galaxy g2.37e12 according to Bondi-Hoyle accretion with AGN FB, gravitational torque accretion without AGN FB, gravitational torque accretion with AGN FB, viscous disc accretion without FB and viscous disc accretion with FB. Additionally, we include the Eddington accretion rates $\dot{M}_{\rm Edd}$ for comparison. The time binning is constant and equal to 216 Myr. For all models that incorporate FB, the galaxy is successfully quenched with SFRs reaching the resolution lower limit of $\sim 3 \times 10^{-3} \, M_{\odot} \rm yr^{-1}$. For both Bondi-Hoyle and the viscous disc accretion models with FB, quenching is complete by $t \sim 8$ Gyrs, while for the gravitational torque model with FB, this occurs later at $t \sim 10$ Gyrs. The gravitational torque and viscous disc models without FB both show similar SFRs and $\dot{M}_{\rm BH}$ as a function of time. The accretion rates show high variability with a $1-\sigma$ scatter of 0.5, $2.5-3$, 1.5 for the Bondi-Hoyle, gravitational torque and the viscous disc models respectively. The SFRs show a minimal scatter of $\sim \pm 0.1-0.2$ dex for all the models.}
    \label{fig:1}
\end{figure}

\begin{figure}[H]
    \centering
    \includegraphics[width = 0.45\textwidth]{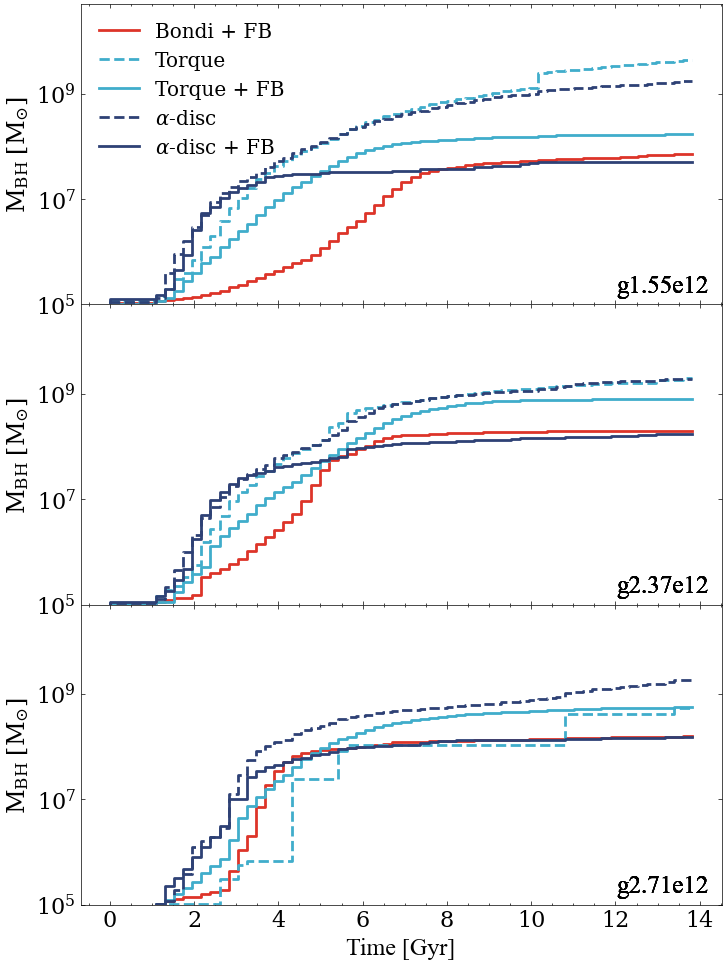}
    \caption{The mass of central black hole (BH) as a function of time for 3 NIHAO galaxies, g1.55e12, g2.37e12 and g2.71e12 for each of the following models: Bondi-Hoyle accretion with AGN FB, gravitational torque accretion without AGN FB, gravitational torque accretion with AGN FB, viscous disc accretion without FB and viscous disc accretion with FB. The time binning is constant and equal to 216 Myr. As expected, runs without AGN FB reach higher BH masses at $z=0$ than corresponding runs with FB. The viscous disc model shows the steepest initial rise, followed by the gravitational torque, and then the Bondi-Hoyle model. Nonetheless, the viscous disc and Bondi-Hoyle models plateau at similar BH masses, while the gravitational torque model reaches higher final BH masses.}
    \label{fig:2}
\end{figure}

As outlined above, the three models under consideration have varying dependencies on the galaxy's interactions and merger history. Bondi-Hoyle accretion shows a strong dependence on the BH mass ($\dot{M}_{\text{BH}} \propto M_{\text{BH}}^2$), while the gravitational torque and viscous disc model show weaker dependencies on the BH mass but stronger dependencies on the BH's local environment. To examine the behaviour of these models, we show the temporal evolution of the central BH's accretion rates and host galaxy star formation rate (SFR) in Figure \ref{fig:1}, and the central BH mass in Figure \ref{fig:2}.  
Figure \ref{fig:1} shows the star formation and BH accretion rates as a function of time for a sample NIHAO galaxy (g2.37e12), for Bondi-Hoyle accretion with AGN FB, and gravitational torque and viscous disc accretion with and without AGN FB. For Bondi-Hoyle accretion, the BH accretion rate increases rapidly, as expected given its prescribed $\dot{M}_{\rm BH} \propto M_{\rm BH}^2$. Initially, the accretion rate is sub-Eddington, but as the BH mass grows, it approaches the Eddington limit and accretes at the maximum rate for approximately one Gyr. Throughout the BH's accretion history, the accretion rate displays a scatter of approximately $\pm 0.5$ dex at the 1-$\sigma$ level. Specifically, between $t \sim 5-7 \, \text{Gyrs}$, there are instances where the instantaneous accretion rate reaches relatively high values of $\dot{M}_{\rm BH} \sim 1 \, M_{\odot} \text{yr}^{-1}$. This heightened accretion will then trigger stronger FB. Consequently, the gas particles surrounding the BH are either expelled by FB or depleted through accretion, leading to a gradual decline in the accretion rate over time. The reduction in gas supply, particularly in the dense central region where star formation is most active, results in a sharp decrease in the SFR. Eventually, the SFR reaches the lower limit of our simulation resolution, approximately $3 \times 10^{-3} \, M_{\odot} \text{yr}^{-1}$. It is worth noting that the SFR exhibits a scatter of approximately $\pm 0.2$ dex on average, across all models considered.

In contrast to the Bondi-Hoyle accretion with AGN FB, the simulation incorporating gravitational torque without AGN FB exhibits substantial BH accretion rates of approximately $\dot{M}_{\rm BH} \sim 10^{-2} \, M_{\odot} \text{yr}^{-1}$ during the early epochs, sustaining Eddington-limited accretion for 1 Gyr. This behaviour aligns with previous findings reported by \citet{angles2013black}. As time progresses, limited only by the gas in-fall rate, the accretion rate increases reaching saturation at $t \sim 7$ Gyrs. As $t \rightarrow 13.8$ Gyrs, we observe that both the BH accretion rate and SFR reach a plateau, with no significant decline observed in either rate. 

On the other hand, when FB is activated, the accretion rate is slower than without FB only sustaining Eddington rates for $\sim$ 200 Myrs. The accretion rate gradually increases to its maximum at $t \sim 7$ Gyrs at $\sim \: 10 \; M_{\odot} \, \text{yr}^{-1}$, and gradually decreases at later times. As the gravitational torque model depends on the local values of parameters that could be highly variable, we find that the 1-$\sigma$ scatter in the accretion rates for the FB and no FB case is large ($\sim \pm 2.5-3$ dex), but can reach smaller values when the accretion rate is high.  

Further, we note a more gradual reduction in the SFR compared to the Bondi-Hoyle model, reaching the minimum SFR at 10 Gyrs. This slower star formation quenching effect for the gravitational torque with the FB run versus the Bondi-Hoyle run might initially seem inconsistent with the higher accretion rates we report. However, Figure \ref{fig:1} shows the time-step averaged (over 216 Myrs) and not the instantaneous accretion rates which directly influence the FB strength. It has been established in the literature that powerful FB episodes are most effective in suppressing star formation, rather than solely considering the total energy released into the system \citep{Schaye2015, fb, horizonbursty, illustris}. Upon examining the instantaneous accretion rates for this BH, we find that the maximum accretion rates are roughly an order of magnitude smaller compared to the Bondi-Hoyle model. This leads to slower heating and depletion of the central gas reservoir. Therefore, within this accretion and FB framework, this results in a longer overall timescale for star formation quenching. Despite the slower decline, FB still plays a crucial role in regulating both star formation and BH growth, as they are reduced when compared to the case without FB.

For the viscous disc model, the accretion remains Eddington-limited for a duration of $\sim $ 2 Gyr after the formation of the BH and maintains a relatively constant level until $t \sim 13.8$ Gyrs. In both the FB and no FB models, the accretion rate exhibits a scatter of approximately $\pm 1.5$ dex at the 1-$\sigma$ level. When FB is included in the model, the BH sustains Eddington-limited accretion for a shorter duration, approximately a few hundred Myrs. During this period, the FB mechanism supplies the surrounding gas with sufficient energy to provide a steady outflow, which results in an earlier onset of star formation quenching compared to the other models. Following this period, the BH exhibits a pattern of BH accretion and star formation quenching, characterized by distinct phases of activity and quiescence. Similar to the Bondi-Hoyle model, the SFR is quenched at $t \sim 7$ Gyrs and falls to its minimum value soon thereafter. 

Similar to the findings reported in \cite{debuhr2011} for the viscous disc model, we find that for all models under investigation, FB does not effectively quench star formation at early times ($t < 4 \, \text{Gyrs}$). However, our FB model differs from that of \citet{debuhr2011} in several aspects. They employ a momentum-based FB approach, while we utilise thermal FB. Additionally, they couple a smaller fraction of the AGN luminosity to the gas and distribute the energy within the accretion radius, whereas we couple the luminosity to the closest 50 gas particles to the BHs. As a result, while our accretion models are similar, the different FB models make it challenging to directly compare the effectiveness of FB in regulating star formation across the different simulations.

In Figure \ref{fig:2}, we show the temporal evolution of the central BH mass. Consistent with expectations, simulations without AGN FB exhibit higher BH masses at redshift zero than their counterparts with FB. This underscores the role of AGN FB in regulating the final BH mass. We found that in the FB runs, the BH mass at $z=0$ is similar for Bondi-Hoyle and viscous disc models, whereas the gravitational torque model produces BHs with higher masses. This discrepancy can be attributed to the weaker instantaneous FB injection with this model, which is less effective in evacuating the gas reservoir. Nevertheless, it is possible to recalibrate the model's parameters to achieve a comparable BH mass to that of the other two models. Additionally, we note that models without FB result in BHs with similar final masses (except for variations that occur when galaxies undergo different merger histories). This suggests that without regulation from FB, the mass of the BH is limited by the total gas budget within the central region of the galaxy.

In the case of galaxy g2.71e12, we find that the BH in the runs with FB grows to a higher mass compared to its no FB counterpart. Upon closer examination, we observe a notably higher stellar density in the inner region of this galaxy. One might expect that stellar FB could hinder BH growth by expelling gas from the vicinity of the BH. However, it is important to note that this effect is not consistently observed across our galaxy sample. Therefore, we exercise caution in drawing definitive conclusions and attribute this variation to the complex and non-linear interactions between AGN FB, BH accretion, and star formation processes. Moreover, we would like to emphasise that the no FB run exhibits a distinct merger history for the BH, which can have various implications for its evolutionary trajectory. These mergers may play a role in shaping the growth and dynamics of the BH, potentially leading to differences in stellar density and other properties within the galaxy.

\begin{figure}[H]
    \centering
    \includegraphics[width = 0.45\textwidth]{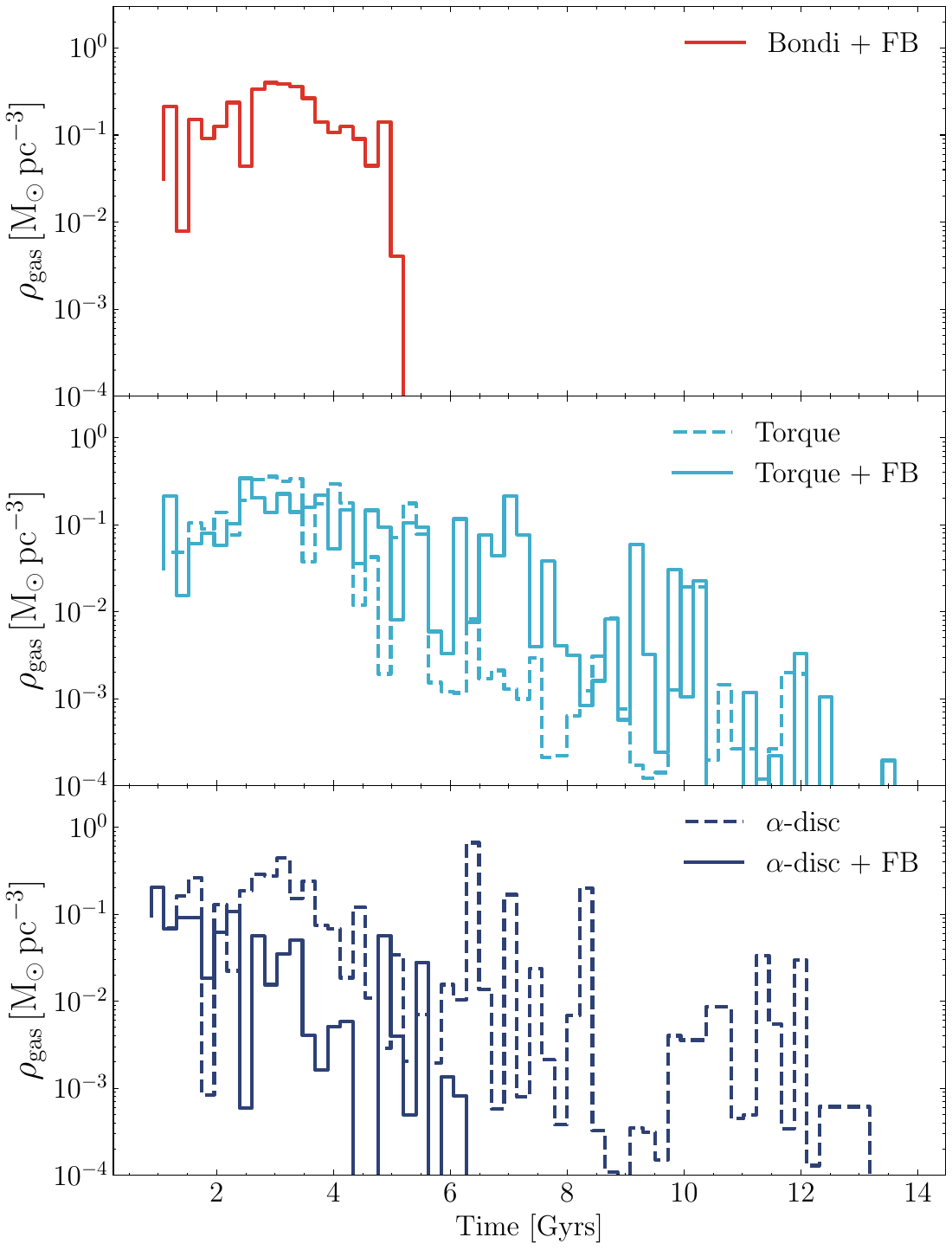}
    \caption{Mean gas density with 1 kpc from the central BH in galaxy g2.37e12 for the Bondi-Hoyle, gravitational torque, and viscous disc accretion models with and without AGN FB. The gas density shows a 1-$\sigma$ scatter of approximately $ \pm (0.1-0.2)$ dex for all models. In the Bondi-Hoyle and viscous disc models with FB, the gas is quickly cleared, while the gravitational torque model exhibits a longer timescale for the density to decrease. The gravitational torque model with FB displays a higher gas density compared to its no FB counterpart, likely due to the gas being utilized through star formation processes.}
    \label{fig:rho}
\end{figure}

Figure \ref{fig:rho} presents the mean gas density, spatially averaged within a 1 kpc radius around the central BH in galaxy g2.37e12, at a given time. This allows us to investigate the efficacy of AGN feedback in clearing out the gas near the BH. Although the mean temperature could serve as a diagnostic for FB efficiency, its variability and susceptibility to noise at our current resolution make it less reliable. Instead, we analyse the gas density, considering that FB heats the gas, causing it to expand and reducing its density in the vicinity of the BH. The measurements reveal a 1$-\sigma$ scatter of approximately $\pm (0.1-0.2)$ dex, indicating relatively minimal variability in the gas distribution at a given time within the 1 kpc region. When FB is implemented in the Bondi-Hoyle and viscous disc models, the gas density experiences a rapid decline, suggesting efficient gas clearing by the FB mechanism relative to the runs without FB. On the contrary, the gravitational torque model with FB exhibits a longer timescale for a significant decrease in gas density. This finding supports that the FB injection for this model requires more time to clear out the gas effectively compared to the other models, consistent with the slower quenching of SFRs highlighted earlier. Notably, when FB is introduced in the gravitational torque model, a higher gas density is observed compared to its no FB counterpart at $t>6$ Gyrs. This phenomenon can potentially be attributed to the gas being used through star formation processes in the case without FB, resulting in a relatively higher gas density in the presence of FB.

Given the steep decline in gas densities over time, it is evident that the accretion likely reaches the accretion radius upper limit before encountering the required 256 or 50 nearest gas particles for the gravitational torque and viscous disc models, respectively. As the number of gas particles used to estimate the accretion rates decreases, this may lead to rate estimates being dominated by noise. In particular, for the gravitational torque model, this would impact the stability of the disc decomposition scheme, which is compromised when employed with a small number of particles during computation. Consequently, this aspect serves as a potential limitation of the model, especially when the resolution near the BH does not ensure sufficient particle sampling.

Furthermore, we compare the results presented in Figure \ref{fig:rho} with those shown in Figure \ref{fig:1}. This comparison reveals a clear correlation between the decline in gas density and the reduction in accretion rates across most of the considered models. For the Bondi-Hoyle model, the gas density exhibits a steep decline, but the decrease in accretion rates is less pronounced. This disparity is attributed to the accretion rate estimation being within a radius encompassing 50 gas particles, which can exceed 1 kpc. In contrast, the accretion rate for the gravitational torque model demonstrates a weaker, implicit dependence on gas density, with other parameters, such as the total mass of baryons within the galactic disc, playing a dominant role. Consequently, similar gas densities in the Torque and Torque + FB model runs yield different accretion rates. Regarding the viscous disc model, a small amount of gas persists after $\sim 6$ Gyrs. Despite this, we observe some non-negligible accretion rates. These significant jumps in accretion rates are likely attributed to rapid or instantaneous inflows of hot gas (not captured within the snapshot times shown in Figure \ref{fig:rho}) within the accretion radius. This gas possesses a high sound speed, facilitating efficient accretion even when the gas surface density is relatively low. Additionally, we note that $\Omega$ is relatively constant on Gyr timescales, exhibiting a gradual, albeit weak, increase. Considering that the accretion rate in the alpha-disc + FB model scales with $\dot{M}_{\rm BH} \propto c_{\rm s}^2 \Sigma/\Omega$, the combination of these factors leads to instantaneous, non-negligible accretion rates.

\begin{figure}[H]
    \centering
    \includegraphics[width = 0.45\textwidth]{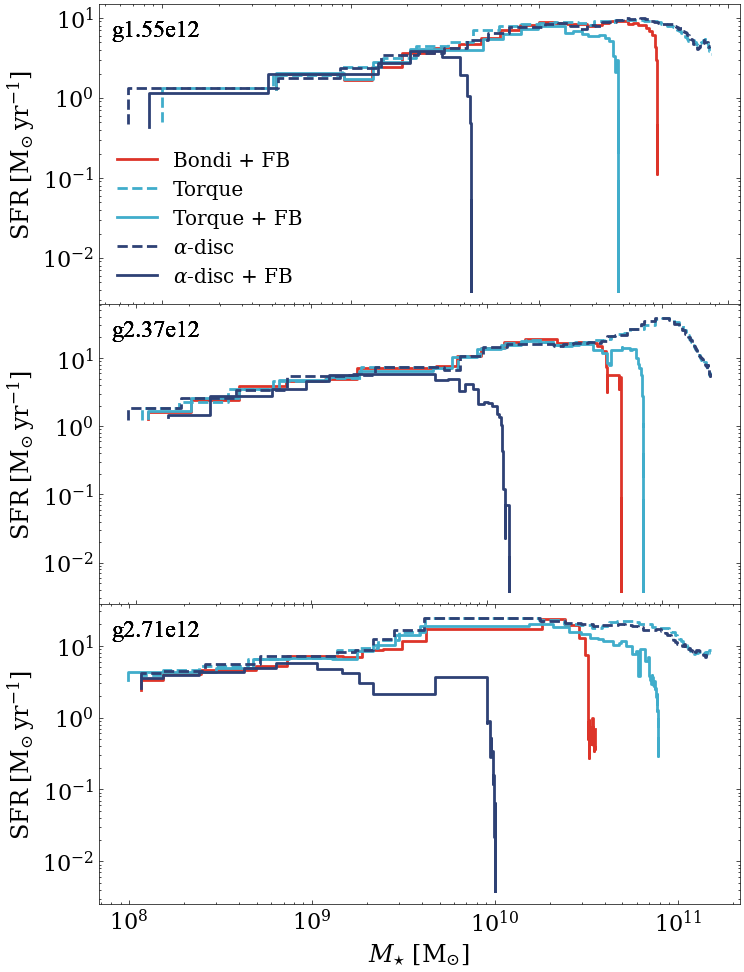}
    \caption{Star formation as a function of stellar mass for three NIHAO galaxies, g1.55e12, g2.37e12 and g2.71e12, according to Bondi-Hoyle accretion with AGN FB, gravitational torque accretion without AGN FB, gravitational torque accretion with AGN FB, viscous disc accretion without FB and viscous disc accretion with FB. The time binning is constant and equal to 216 Myr. For runs with AGN FB, star formation is reduced by several orders of magnitude relative to their no FB equivalent runs, often reaching the resolution lower limit of $3\times10^{-3}\, M_{\odot} \text{yr}^{-1}$. With our current FB prescription, the viscous disc model with FB produces galaxies with a much lower stellar mass than other models.}
    \label{fig:3}
\end{figure}

To investigate the effect of AGN FB on SFR, we show the temporal evolution of the SFR as a function of stellar mass for three NIHAO galaxies, g1.55e12, g2.37e12, and g1.71e12 in Figure \ref{fig:3}, and compare the SFR evolution for each model with and without FB. We observe little to no variation in the effectiveness of quenching SFR across the different models that do not incorporate FB. In all cases, the SFR halts when all cool dense gas is converted into stars. However, simulations that incorporate FB regularly show a more substantial decline in SFR across all models, with the viscous disc model consistently quenching star formation at lower stellar masses than the Bondi-Hoyle or gravitational torque models. This arises as our FB normalization was calibrated using the Bondi-Hoyle simulations. As these simulations predict low BH accretion rates at early times, and thus weaker FB, they provide more time for star formation to occur. Consequently, to produce galaxies with realistic stellar masses, lower FB efficiency values would be needed for models with high initial accretion rates such as the viscous disc model.

\begin{figure}[H]
    \centering
    \includegraphics[width = 0.45\textwidth]{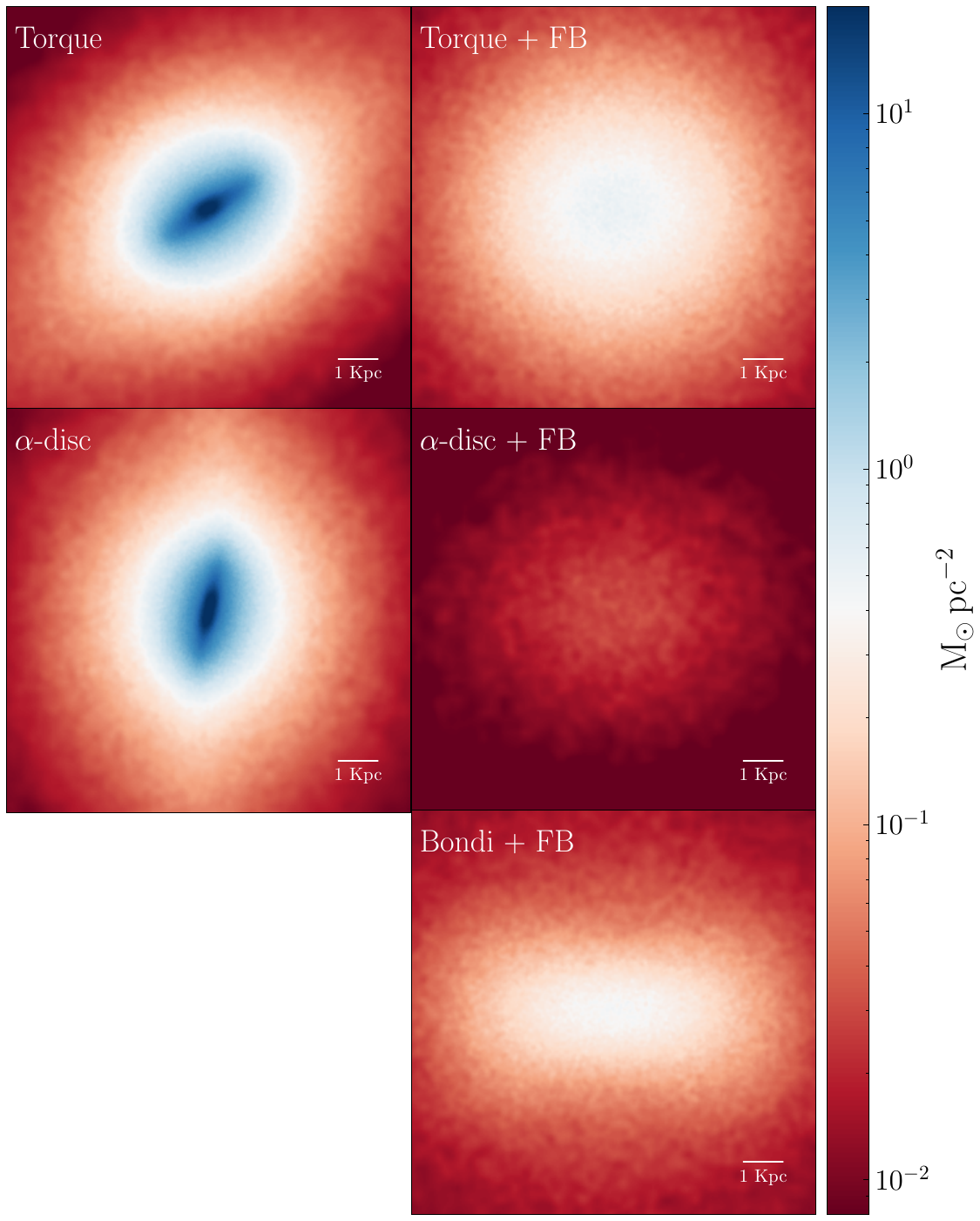}
    \caption{Face-on projection of the stellar surface density maps for galaxy g2.37e12 with (right) and without (left) AGN FB for gravitational torque, viscous disc, and Bondi-Hoyle accretion models. In the absence of AGN FB, the stars exhibit high compactness and are concentrated predominantly at the centre of the galaxy. When FB is activated, star formation is suppressed at the galaxy's centre, resulting in central stellar densities that are approximately one order of magnitude lower or approximately two times for the viscous disc model.}
    \label{fig:morph}
\end{figure}

In addition to its impact on the overall SFR, the choice of accretion model significantly influences the morphology of galaxies. Figure \ref{fig:morph} illustrates the face-on stellar surface density of galaxies g2.37e12 for the gravitational torque, viscous disc, and Bondi-Hoyle accretion models, both with and without FB. In the absence of FB, the galaxies appear nearly identical, with a high and concentrated stellar surface density at the centre. This similarity is expected, as shown in Figure \ref{fig:1}, where these models exhibit similar BH accretion and SFR histories. However, when FB is enabled, noticeable differences emerge in the galaxy morphology. Comparing the FB runs to their counterparts without FB, we observe that FB leads to galaxies with a less compact central stellar surface density, which is approximately one order of magnitude less dense. It is important to note that in the case of the viscous disc model, the current FB prescription results in galaxies that are overly quenched, exhibiting a surface density of stars roughly two orders of magnitude lower than the no FB counterpart. Overall, the combined effect of the choice of accretion model when coupled with FB impacts both the timing and spatial distribution of star formation, thereby shaping the overall morphology of the galaxy.

\subsection{Evolution of Scaling Relations}
\begin{figure}[H]
    \centering
    \includegraphics[width = 0.45\textwidth]{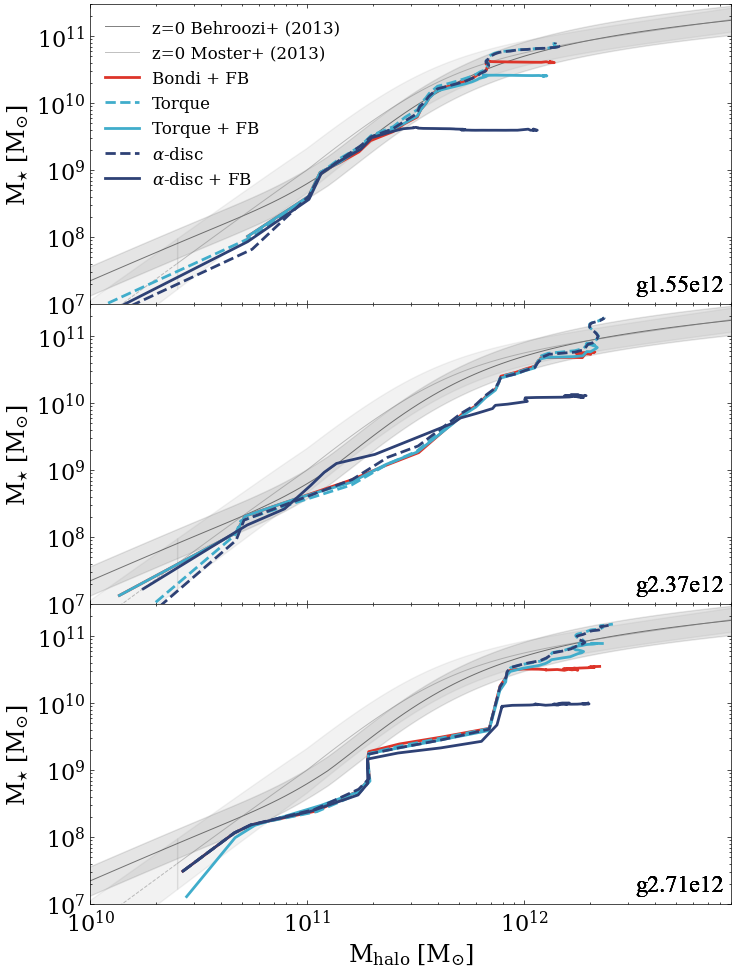}
    \caption{Evolutionary tracks for 3 NIHAO galaxies, g1.55e12, g2.37e12 and g2.71e12 across the Stellar Mass-Halo Mass plane for each of the following models: Bondi-Hoyle accretion with AGN FB, gravitational torque accretion without AGN FB, gravitational torque accretion with AGN FB, viscous disc accretion without FB and viscous disc accretion with FB. The dark grey and light grey solid lines correspond to the observed correlations reported by \citet{behroozi_etal13} and \citet{moster_etal13} respectively with a scatter of $\pm 0.28$ dex. All models, except the viscous disc model, show a relatively consistent agreement with the observed relations. This indicates that further tuning of the FB parameters for the viscous disc model would be required in the future.}
    \label{fig:4}
\end{figure}

In Figure \ref{fig:4}, we show the temporal evolution for the three NIHAO galaxies across the Stellar Mass-Halo Mass plane. We note that for each galaxy, all models generally align with the observed relation at $z=0$, except for the viscous disc model with FB runs which undershoot the relation. This behaviour is expected given the results presented in earlier figures, where the viscous disc model under-produces stars.

\begin{figure}[H]
    \centering
    \includegraphics[width = 0.45\textwidth]{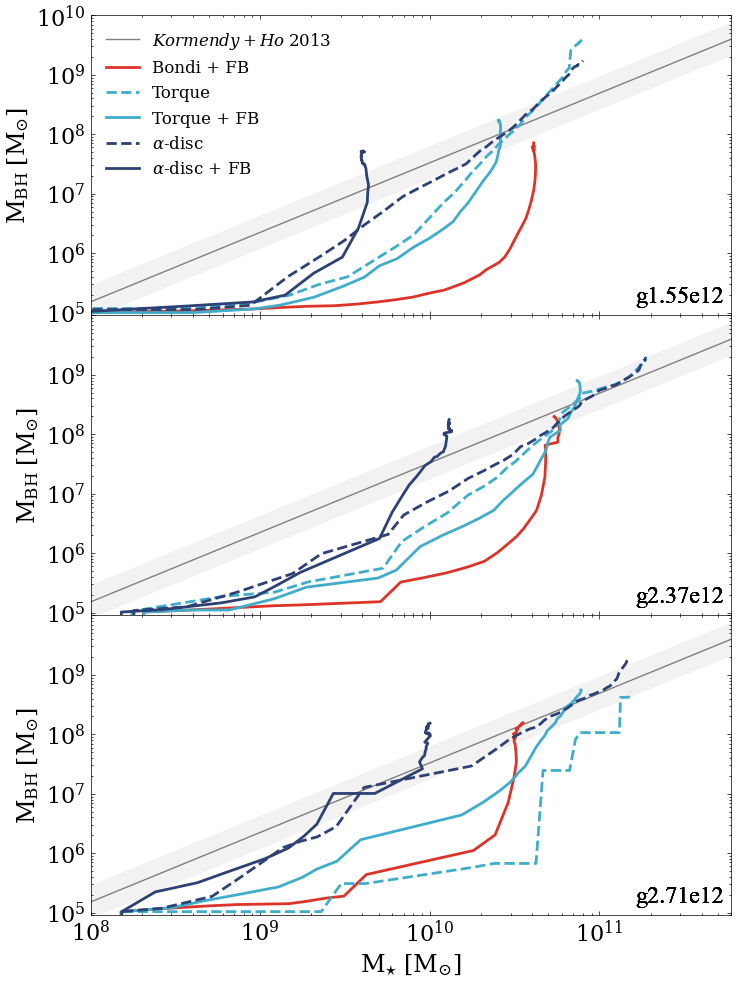}
    \caption{Evolutionary tracks for 3 NIHAO galaxies g1.55e12, g2.37e12 and g2.71e12 across the Black Hole (BH}) Mass-Stellar Mass plane for each of the following models: Bondi-Hoyle accretion with AGN FB, gravitational torque accretion without AGN FB, gravitational torque accretion with AGN FB, viscous disc accretion without FB and viscous disc accretion with FB. The dark grey solid line corresponds to the observed correlation reported by \citet{kormendy_ho13} with a scatter of $\pm 0.28$ dex. For the Bondi-Hoyle model, the track has a shallow gradient at low stellar masses that transitions into a steep (almost vertical) rise at the higher stellar mass end. The tracks for the gravitational torque model with and without FB end up on the observed relation, however for galaxy g1.55e12 the gravitational torque run without FB overshoots the relation. The viscous disc model with FB results in galaxies that undershoot the BH and stellar masses relative to all other models. The viscous disc model run without FB evolves roughly parallel to the relations showing co-existence between star formation and BH growth. 
    \label{fig:5}
\end{figure}

\begin{figure}[H]
    \centering
    \includegraphics[width = 0.45\textwidth]{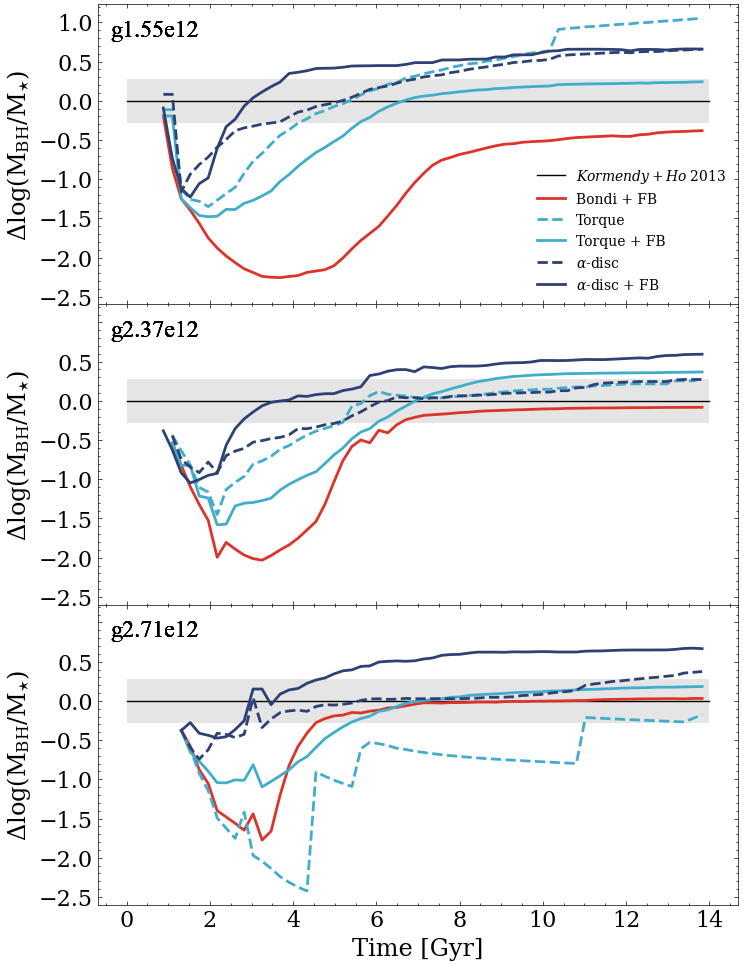}
    \caption{Temporal evolution of the deviation for 3 NIHAO galaxies g1.55e12, g2.37e12 and g2.71e12 from the \citet{kormendy_ho13} observed relation. We calculate $\Delta \text{log}(M_{\rm BH}/M_{\star})$ of the galaxy as the vertical distance of the ratio from the observed relation's ratio in log space. The dark grey solid line corresponds to the observed correlation reported by \citet{kormendy_ho13} with a scatter of $\pm 0.28$ dex. All models show minimal evolution after $z\sim1$ and minimal deviation from the observed relation even for runs without FB. }
    \label{fig:6}
\end{figure}

Figures \ref{fig:5} and \ref{fig:6} show the temporal evolution along the BH Mass-Stellar Mass relation and the vertical deviation from the \cite{kormendy_ho13} relation respectively, for the three NIHAO galaxies mentioned earlier. In the Bondi-Hoyle simulations, initially when the BH mass is low, and therefore the accretion rate ($\dot{M}_{\rm BH} \propto M_{\rm BH}^2$) and FB rate ($\dot{E} \propto \dot{M}_{\rm BH}$) is also low. Consequently, the stellar mass grows at a much faster rate than the BH mass. Given these scalings, as the BH mass increases, the accretion and FB rates rise dramatically.  This results in a distinctive parabolic evolutionary trajectory, which is a characteristic feature of this model \citep{sharma2020black}. However, the tracks are different than those obtained using the Illustris \citep{illustris}, and the Horizon-AGN simulations \citep{volonteri2016cosmic}. This could be attributed to differences in the implementations of FB, FB mechanisms, and simulation resolution across the different projects.

For galaxies g1.55e12 and g2.37e12, the gravitational torque model displays similar evolutionary tracks when run with and without FB. However, the no FB runs result in slightly more massive BHs. These models begin to diverge as $M_\star \rightarrow 10^{10} M_{\odot}$, where the runs with FB have slowed down BH growth and quenched star formation. Meanwhile, as there is no FB limiting the growth of the BH in the runs without FB, both the BH and stellar masses continue to increase. As previously mentioned, galaxy g2.71e12 exhibits a distinct behaviour where the absence of AGN FB results in slower growth of the BH compared to the stellar mass. In this case, star formation becomes the primary process responsible for gas consumption, utilizing a larger fraction of the available gas. Furthermore, due to the divergent merger history of the galaxy, characterised by significant jumps in the figure, it becomes challenging to accurately isolate and evaluate the direct impact of AGN FB. We compare our no FB runs to those shown in \citet{angles2016} using the FIRE project, and find qualitatively similar tracks. In general, our findings align with those of \cite{angles2013black}, suggesting that the growth of BH does not necessarily require regulation through FB to match the scaling relations.

The viscous disc model with FB simulation exhibits an initial slow rise, followed by a steep increase in the BH mass. This is accompanied by a low rate of stellar formation as demonstrated in Figures \ref{fig:2} and \ref{fig:3} respectively. And while at $z=0$, the BH and stellar masses lie slightly above the relation, they undershoot both the stellar and BH mass predicted by the other models at the current FB calibration. When the model is used without FB, the BH mass and stellar mass gradually increase following a trajectory that is roughly parallel to the relation until $z=0$. This suggests that for this model, it is likely that the BH and stellar masses co-exist regulating their growth by competing for a common gas supply. It is worth noting that our conclusion differs from that of \cite{debuhr2011}, where they argue that FB is necessary to regulate BH growth. However, comparing our results to theirs is challenging due to several factors. Firstly, they resolve accretion on smaller scales with higher resolution. Additionally, they adopted a different FB prescription that is momentum-coupled rather than thermally coupled to the gas.

To distinguish the evolution of the models under consideration, we show their deviations from the observed relation as a function of time in Figure \ref{fig:6}. In all models, slight deviations are observed from the expected relation from $z\sim 1 \to 0$, typically within the range $\pm 0.5$ dex. The models diverge at higher redshifts, where observational constraints are limited. Thus, it is difficult to conclude which model most accurately describes the evolution of BHs in their host galaxies. Further, as models without FB match the observed relations at $z=0$, and show co-evolution along the observed relations, 
this suggests that BH FB may not be the sole mechanism responsible for driving the tight correlations between the BH mass and the galaxy's stellar properties. Our simulations without FB demonstrate that the observed relations naturally emerge as both the BH and the stars compete for a shared gas reservoir, and thus co-evolve with each other. This does not imply that FB does not have any role in shaping the co-evolution but may imply that it has a secondary role, on the contrary of what is usually assumed \citep[see][for an extensive review]{kormendy_ho13}. 


\subsection{Parameter Study: Accretion Radius}
For the simulations presented above, we evaluate the maximum accretion rate at an accretion radius of $R_0 \sim$1 kpc. The accretion radius is particularly interesting as it affects the accretion rate explicitly as shown in Equation \ref{eq3} for the gravitational torque model ($\dot{M}_{\rm Torque} \propto R_0^{-3/2}$), and Equation \ref{eq6} for the viscous disc model ($\dot{M}_{\alpha\text{-disc}} \propto R_0^{-1/2}$). It also affects the accretion rate implicitly by changing the other parameters in the equations that are evaluated at $R_0$. In this section, we investigate the effect of varying the accretion radius on our models and present the results for different values of R0 (1 kpc, 0.5 kpc, and 0.25 kpc) for the NIHAO galaxy g6.86e12. We omit the results for the gravitational torque model with $R_0 = 0.25 \, \rm kpc$ due to the model's instability at the current resolution of our simulations. To probe the direct effect of the accretion radius on the accretion rates, we eliminate BH FB for the simulations presented in this section. 

\subsubsection{Effect on Accretion Rate}
\begin{figure}[H]
    \centering
    \includegraphics[width = 0.45\textwidth]{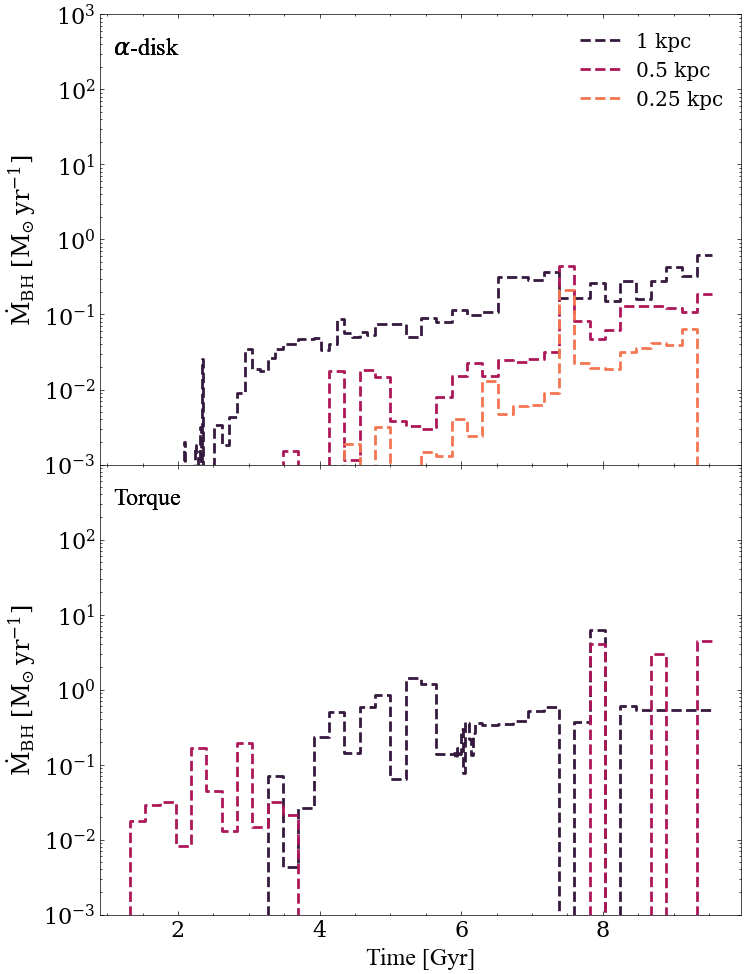}
    \caption{Black hole (BH)} accretion rates of the central BH as a function of time for galaxy g6.86e12 according to viscous disc accretion (top) for $R_0 = 1 \, \rm kpc$, $R_0 = 0.5 \, \rm kpc$ and $R_0 = 0.25 \, \rm kpc$, and gravitational torque accretion (bottom) without FB for $R_0 = 1 \, \rm kpc$ and $R_0 = 0.5\, \rm kpc$. As expected, decreasing $R_0$ decreases BH accretion rates. We exclude the plot for the gravitational torque model with $R_0=0.25 \, \rm kpc$ as it was numerically unstable at our current resolution. 
    \label{fig:radius_rates}
\end{figure}
In Figure \ref{fig:radius_rates}, we show the BH accretion rate for viscous disc accretion (top) for $R_0 = 1 \, \rm kpc$, $R_0 = 0.5 \, \rm kpc$ and $R_0 = 0.25 \, \rm kpc$, and gravitational torque accretion (bottom) with accretion radii of $R_0 = 1 \, \rm kpc$ and $R_0 = 0.5 \, \rm kpc$.  Upon analysis, we find that the primary factor driving the higher accretion rate for larger radii is the gas mass $M_{\rm gas}(R_0)$ incorporated in the gas surface density $\Sigma$. This observation supports our initial expectation and is consistent with the findings reported in \citet{debuhr2011}. 
 
For the gravitational torque model, we find that the $R_0 =0.5$ BH undergoes bursts of high accretion rates that are brief relative to the $R_0=1 \, \rm kpc$ run resulting in an overall lower BH mass. This burstiness is likely a consequence of the strong dependence of this model on the local environment surrounding the BH. However, it is important to note that evaluating these parameters within radii smaller than 1 kpc, given our restricted resolution, introduces significant noise and uncertainties. The limited resolution at smaller accretion radii restricts our ability to accurately determine the properties and behaviour of the gas. The kinematic decomposition scheme employed in our study requires a larger number of particles for a well-behaved decomposition. As the size of the accretion disc decreases, the number of particles within the region of interest also decreases, leading to a less reliable decomposition of the gas kinematics. Therefore, at our current resolution, accurately capturing the intended behaviour of the gravitational torque model proves challenging. However, when a sufficient number of gas particles are available for the decomposition, we find that the accretion rate is primarily determined by the mass of gas in the disc, which aligns with the findings reported by\cite{angles2013black}. 

\subsubsection{Effect on Scaling Relations}
\begin{figure}[H]
    \centering
    \includegraphics[width = 0.45\textwidth]{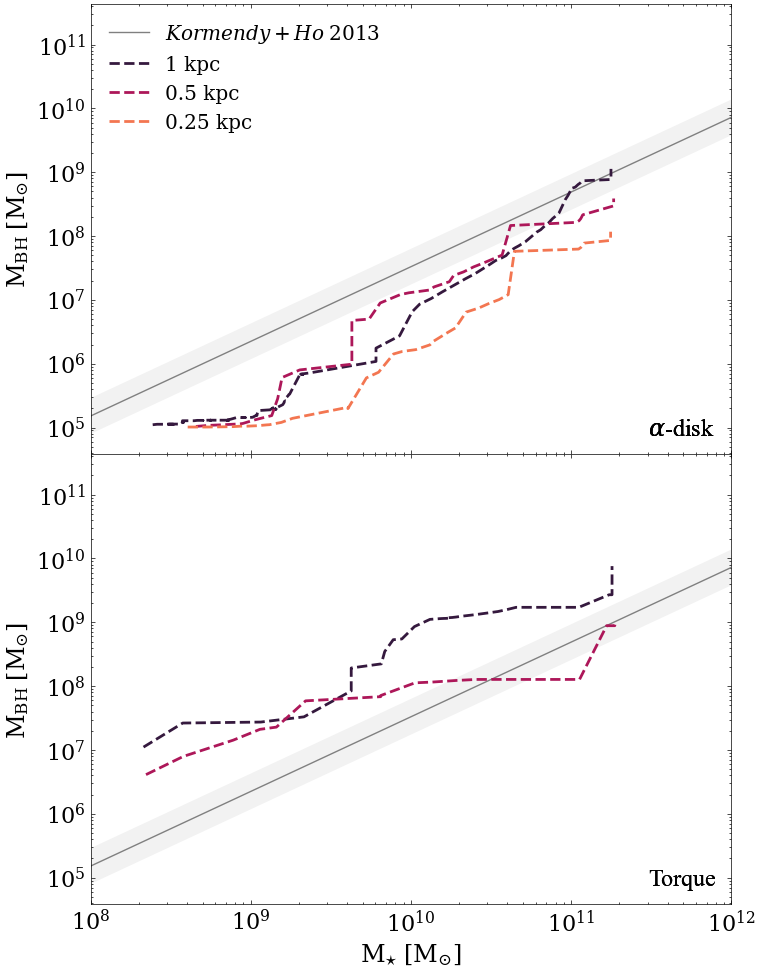}
    \caption{Evolutionary tracks for galaxy g6.86e12 across the Black Hole (BH) Mass-Stellar Mass plane for viscous disc accretion (top) and gravitational torque accretion (bottom) without AGN FB for $R_0 = 1 \, \rm kpc$, $R_0 = 0.5 \, \rm kpc$ and $R_0 = 0.25 \, \rm kpc$. The dark grey solid line corresponds to the observed correlation reported by \citet{kormendy_ho13} with a scatter of $ \pm 0.28$ dex. For both models, reducing the value of $R_0$ leads to changes only in the normalization of the final BH mass without significantly altering the slope of the track.} 
    \label{fig:mbh_mstar_r}
\end{figure}

As we do not include BH FB in these simulations and the BH accretion rate is not sufficiently high to deplete the gas within the central region of the galaxy, varying the accretion radius has no effect on the SFRs across all our runs. This is shown in Figure \ref{fig:mbh_mstar_r} where we show the evolution of the $M_{\rm BH}-M_{\star}$ relation up to redshift 1 for runs with different accretion radii. We find that all runs reach the same stellar mass and have comparable slopes across the $M_{\rm BH}-M_{\star}$ plane. Specifically, for the viscous disc model, we deduce that the choice of accretion radius does not change the slope along the BH Mass-Stellar Mass plane, but merely affects the normalisation of the evolution curve. We note similar trends for the gravitational torque model. Hence, we infer that our results are stable against the choice of accretion radius. However, to achieve similar results, re-calibration would be required when changing the accretion radius or the resolution of the simulations. 

\section{Conclusions}
\label{sec:conclusions}
    Tight correlations between massive BHs and their host galaxy's stellar properties such as stellar mass and stellar dispersion velocity have been found through extensive observations. These correlations might suggest that processes that occur on small scales governing the growth of BHs are also coupled to galaxy-wide processes that extend to much larger spatial and mass scales. The mechanism that drives this possible coupling remains poorly understood which motivates the following question: Does the mass of the BH co-evolve with the stellar mass due to regulation through AGN FB, or does the shared gaseous environment result in their co-existence? 
    
    Most of the research currently focuses on studying the earlier question by examining different prescriptions of AGN FB for simulations of galaxy formation. The Bondi-Hoyle prescription suggests that FB is necessary to regulate BH growth, although it is unclear if the resolution of the simulations accurately captures the accretion rate where this model is applicable. Other models, such as the viscous disc model proposed by \cite{debuhr2011}, also indicate the need for FB to produce BHs with realistic masses. On the other hand, the gravitational torque-based accretion model proposed by \citet{hopkins2011analytic} suggests that explicit regulation through FB processes may not be necessary to reproduce the observed scaling relations. 
    
    In this paper, we emphasise the significant impact of the choice of the BH accretion model on the origin and evolution of the scaling relations which should not be overlooked. In this work, we studied the evolution of the scaling relations by running simulations for a set of four NIHAO galaxies according to three BH accretion models, the Bondi-Hoyle, gravitational torque, and viscous disc models.  We find that the choice of the BH accretion model is critical to answering the question above as it changes how galaxies evolve regardless of FB. 
    
    While taking BH FB into account for galaxy formation simulations is important to set the final mass of the central SMBH and regulate star formation, our study suggests that FB might not be the sole mechanism giving rise to the observed scaling relations.
   We find that both the gravitational torque and viscous disc models evolve following the $M_{\text{BH}}-M_{\star}$ relation observed at $z=0$, without the need for any BH FB. This suggests that, in these models, BHs co-exist with the host galaxy's stellar mass as their growth is regulated through a common gas supply within the central bulge. When AGN FB is incorporated into the models, it plays a vital role in regulating the gas supply for both BH accretion and star formation processes. This regulation ultimately determines the final masses of both the BH and the stellar component. Therefore, it is essential to accurately model both the accretion and FB processes, as they are likely to both contribute to the emergence of the scaling relations. In other words, star formation and BH accretion co-exist within a shared gas reservoir, which determines their respective rates. However, the presence of AGN FB impacts the total gas budget, influencing the final masses of both the BH and the stellar component.
    
    Further, we check that our results are stable against our choice of the accretion radius $R_0$ in both models. We find that varying $R_0$ merely affects the overall normalization of the evolutionary tracks along the $M_{\text{BH}}-M_{\star}$ relation, which could be mitigated by a re-calibration of the FB efficiency $\epsilon_f$ from Equation \ref{eq:fb}. 
    
    Our study suggests that the debate between co-evolution vs. co-existence is still to be settled, and that, as usual, the devil is in the details, when we try to model in cosmological simulations the complex physics of BH accretion and FB.

\section*{Acknowledgements}
 This research was carried out using the high performance computing resources at New York University Abu Dhabi. We use the PYNBODY software package PYNBODY \citep{pynbody} for our analyses. 

\section*{Data availability}
The data underlying this article will be shared on reasonable
request to the corresponding author.

\bibliographystyle{mnras}
\bibliography{ref1} 
\bsp	
\label{lastpage}
\end{document}